\documentclass[times,sort&compress,3p]{elsarticle}
\journal{Journal of Multivariate Analysis}
\usepackage[labelfont=bf]{caption}

\usepackage{amsmath,amsfonts,amssymb,amsthm,booktabs,color,epsfig,graphicx,hyperref,url, multicol, multirow}
\allowdisplaybreaks

\theoremstyle{plain}
\newtheorem{theorem}{Theorem}

\theoremstyle{definition}

\newcommand{\vect}[1]{\boldsymbol{#1}}

\begin{document}

\begin{frontmatter}

\title{\texttt{fastMI}: a fast and consistent copula-based nonparametric estimator of mutual information}

\author[1]{Soumik Purkayastha}
\author[1]{Peter X.-K. Song\corref{mycorrespondingauthor}}

\address[1]{Department of Biostatistics, University of Michigan, Ann Arbor, MI 48109, USA}

\cortext[mycorrespondingauthor]{Corresponding author. Email address: \url{pxsong@umich.edu}}

\begin{abstract}
As a fundamental concept in information theory, mutual information ($MI$) has been commonly applied to quantify association between random vectors. Most existing nonparametric estimators of $MI$ have unstable statistical performance since they involve parameter tuning. We develop a consistent and powerful estimator, called \texttt{fastMI}, that does not incur any parameter tuning. Based on a copula formulation, \texttt{fastMI} estimates $MI$ by leveraging Fast Fourier transform-based estimation of the underlying density. Extensive simulation studies reveal that \texttt{fastMI} outperforms state-of-the-art estimators with improved estimation accuracy and reduced run time for large data sets. \texttt{fastMI} provides a powerful test for independence that exhibits satisfactory type I error control. Anticipating that it will be a powerful tool in estimating mutual information in a broad range of data, we develop an \textsf{R} package \texttt{fastMI} for broader dissemination. 
\end{abstract}


\end{frontmatter}

\section{Introduction\label{sec:1}}

Investigating dependence between two random variables is a key issue in statistical science. It is known that classical measures like Pearson's $r$, Kendall's $\tau$, and Spearman's $\rho$ \cite{Everitt_2010}, although widely used in practice, 
are incapable of capturing a non-linear, non-monotonic association, which cannot cannot be properly estimated using ranks and their monotonic transformations. To address this practical need, more sophisticated measures, like the distance correlation (dCor) \cite{Szekely_2007}, the Heller-Heller-Gorfine (HHG) \cite{Heller_2012} statistic, and the maximal information coefficient (MIC) \cite{Reshef2011} have been introduced to study more complex association patterns such as the dependence of two random vectors. 

Among many non-linear association metrics, mutual information ($MI$) \cite{Cover_2005} has recently re-emerged in the statistics and machine learning literature with many exciting applications \cite{Wang_2009}.  Originally introduced in a communication theory context \cite{Shannon_1948}, $MI$ presents a remarkably general and intuitive measure of dependence. Its widespread use in practice is due largely to the self-equitability \cite{Kinney2014} of $MI$ - the ability to characterize dependency strength for both linear and non-linear relationships.

For continuous data, which shall be the focus of this paper, there are three distinct $MI$ estimation approaches. The first approach implements a binning method to group continuous data into different bins and estimates $MI$ from the binned data \cite{Strong_1998, Paninski2003}. The success of this simple method depends heavily on proper specification of both number and position of said bins. Another approach is based on a k-nearest neighbors (kNN) estimation method, utilized by the Kraskov-St\"{o}gbauer-Grassberger (KSG) estimator \cite{Kraskov2004}. As is the case with all kNN-based methods, the KSG estimator greatly depends on properly specifying the number of neighbors. The third approach is based on estimates of probability density functions (PDFs), using histograms, kernel density estimation (KDE) \cite{moon_1995}, B-splines \cite{daub_2004}, or wavelets \cite{peter_2008}. This nonparametric approach typically relies on a tuning parameter (e.g. bandwidth) that needs to be specified in the estimation routine. 

Although the approaches mentioned above demonstrate good  properties \cite{Paninski2003, Kraskov2004, moon_1995, daub_2004, peter_2008}, they are all sensitive to proper specification of tuning parameter(s) in a chosen smoothing technique. As a consequence, the resulting estimators may be numerically unstable and/or suffer from serious bias. A recent study \cite{Zeng2018} presents a tuning-free KDE approach to estimate $MI$ in which the bandwidth parameter is automatically set to maximize the jackknifed version of $MI$ (henceforth referred to as $JMI$). Not only does this method exhibit better estimation efficiency than other existing $MI$ estimation approaches but it also provides a stable hypothesis test for independence that is shown to be more powerful than its competitors such as the dCor, HHG, or MIC. According to existing literature, $JMI$ serves as the current gold standard for estimating $MI$ as well as is the choice of test for independence \cite{Zeng2018}.

Along the line of tuning-free estimation, an improvement on $JMI$ is proposed in this paper. This new estimator is motivated by the self-consistent density estimator proposed by \cite{Bernacchia2011, OBrien2016}  to minimize the mean integrated squared error (MISE) between the estimated density and the true density without incurring any manual parameter tuning. The estimation process relies on fast Fourier transforms (FFT). 

Furthermore, utilizing this  `optimal' density estimator, we propose a plug-in estimator of $MI$, termed as the \texttt{fastMI}, which is shown to be consistent and manyfold faster than the original  $JMI$ for large data sets.  This proposed \texttt{fastMI} automatically determines the bandwidth by minimizing the MISE objective function in a data-adaptive way.

Simulation studies comparing estimation accuracy reveal \texttt{fastMI} outperforms the current gold standard $JMI$ estimator. Through extensive numerical experiments, \texttt{fastMI} demonstrates improved estimation efficiency, higher empirical power when testing for independence, and reduced computation time.  All these lead to a recommendation of our new methodology to practitioners. 

The rest of the paper is organized as follows. In Section \ref{methods} we define $MI$ and underline a key connection of $MI$ with copula. In Section \ref{sec:sc_estimation} we describe the self-consistent density estimation method in detail. In Section \ref{sec:theory} we provide theoretical guarantees on the asymptotic behaviour of the \texttt{fastMI} estimator. We present extensive simulation studies in Section \ref{simulations} to highlight the strengths and advantages of using \texttt{fastMI} over the existing state-of-the-art estimator, the $JMI$ estimator. Further, we benchmark our findings by means of an empirical copula estimator-based $MI$ as well. Finally, we present an application of \texttt{fastMI} to real data in Section \ref{rda} before summarizing our findings in the concluding Section \ref{discussion}. All heavy technical details are included in Section \ref{sec:tech}.

\section{Methods\label{methods}}

 To formalize the problem, suppose that $d \in \mathbb{N}$ can be written as $d = p + q$ for some $p, q \in \mathbb{N}$, that $\vect{X}$ and $\vect{Y}$ are random vectors taking values in $\mathbb{R}^p$ and $\mathbb{R}^q$, respectively, and that $\vect{Z} = (\vect{X}^\top, \vect{Y}^\top)^\top$ has density $f_{\vect{XY}}$ with respect to Lebesgue measure on $\mathbb{R}^d$. We write $f_{\vect{X}}$ and $f_{\vect{Y}}$ for the marginal Lebesgue densities of the random vectors $\vect{X}$ and $\vect{Y}$, respectively. Given independent and identically distributed copies $\left\{ \vect{Z}_1,\ldots, \vect{Z}_n \right\}$ of $\vect{Z}$, one of our primary tasks is to test the null hypothesis $H_0$ that the \textcolor{blue}{vectors} $\vect{X}$ and $\vect{Y}$ are independent against the alternative that $\vect{X}$ and $\vect{Y}$ are not independent.

\subsection{Mutual information and its copula-based formulation}
In order to test the independence of $\vect{X}$ and $\vect{Y}$ we consider mutual information $MI(\vect{X}, \vect{Y})$, defined by
\begin{equation*}
\label{eq:mi_defn}
MI(\vect{X}, \vect{Y}) = \mathbb{E}_{\vect{XY}} \left[ \ln \left\{ \frac{f_{\vect{XY}}(\vect{X}, \vect{Y})}{f_{\vect{X}}(\vect{X}) f_{\vect{Y}}(\vect{Y})} \right\} \right], 
\end{equation*} 
where $\mathbb{E}_{\vect{XY}}$ denotes expectation under the joint density $f_{\vect{XY}}$. It is known that $MI$ is a valid measure of association; that is, it is equal to zero if and only if $\vect{X}$ and $\vect{Y}$ are independent and positive otherwise. Larger values of $MI$ indicate a stronger association. $MI$ is invariant under monotonic transformations, which is an important property allowing various rank-based transformations. Further, $MI$ satisfies the self-equitability condition \cite{Zeng2018}, implying that it detects associations without a bias for specific association patterns, unlike the $MIC$ \cite{Kinney2014}.

Interestingly, $MI$ may be rewritten as a function of copulas, a class of dependence models \citep{Joe_Harry2014-06-26}. Note that by using Sklar's theorem \cite{czado_2019}, we can apply monotonic marginal transformations to reduce the technical complexity by using  uniform transformations $\vect{U_X} = \left(F_{X_1}(X_1), \ldots, F_{X_p}(X_p) \right)^\top$ and $\vect{U_Y} = \left(F_{Y_1}(Y_1), \ldots, F_{Y_q}(Y_q) \right)^\top$, where $F_{X_i}$ and $F_{Y_j}$ are the cumulative distribution functions (CDFs) of $X_i$ and $Y_j$, respectively, for $1 \leq i \leq p$ and $1 \leq j \leq q$. We use $c_{\vect{U_X}\vect{U_Y}}$, $c_{\vect{U_X}}$, and $c_{\vect{U_Y}}$ to denote the copula density functions of $(\vect{U_X}^\top, \vect{U_Y}^\top)^\top$, $\vect{U_X}$, and $\vect{U_Y}$ respectively. Using this marginal uniform transformation trick, we arrive at an alternative copula-based formulation of $MI$:
\begin{align}
\label{eq:copula-entropy}
MI(\vect{X}, \vect{Y})  &= MI(\vect{U_X}, \vect{U_Y}) \notag \\ &= \int_{[0, 1]^p} \int_{[0, 1]^q}  \ln \left\{ \frac{c_{\vect{U_X}\vect{U_Y}}(\vect{u_X}, \vect{u_Y})}{c_{\vect{U_X}}(\vect{u_X})c_{\vect{U_Y}}(\vect{u_Y})} \right\} c_{\vect{U_X}\vect{U_Y}}(\vect{u_X}, \vect{u_Y}) d\vect{u_X} d\vect{u_Y}.
\end{align}
$MI$ in Equation \ref{eq:copula-entropy} may be estimated by plug-in copula density estimators.
The marginal uniform transformation trick allows us to consider ranks instead of raw data, making copula-based estimation methods robust to any marginal irregularity, in contrast with methods which use raw data to estimate $MI$, such as the kNN-based KSG estimator. Recognizing $MI$ as an integral in  Equation \ref{eq:copula-entropy}, we may invoke data generative methods, including classical Monte Carlo methods \cite{robert_2010} for estimation purposes. 
The estimated copula density functions $\hat{c}_{\vect{U_X}\vect{U_Y}}$, $\hat{c}_{\vect{U_X}}$, and $\hat{c}_{\vect{U_Y}}$ are used to obtain an estimate of the underlying $MI$. Using a fast Fourier transform-based density estimation technique \citep{Bernacchia2011, OBrien2016}, which is described in greater detail in Section \ref{sec:np_cop} and Section \ref{sec:sc_estimation}, we obtain said copula density estimates. This estimation routine will be utilized in a powerful  $MI$-based hypothesis test for independence of $\vect{X}$ and $\vect{Y}$.

\subsection{Improved nonparametric estimation of copula density using probit transformation}
\label{sec:np_cop}

From Equation \ref{eq:copula-entropy} we note the need to estimate three multivariate copula density functions, namely ${c}_{\vect{U_X}\vect{U_Y}}$, ${c}_{\vect{U_X}}$, and ${c}_{\vect{U_Y}}$. Here we describe the nonparametric KDE of a generic $d$-variate copula density function for vector $\vect{U} = \left(U_1, \ldots, U_d \right)^\top$ on unit hypercube $[0, 1]^d$. The naive KDE \cite{Silverman1986}, is not suitable for unit hypercube-supported copula densities, mainly because it is known to be heavily affected by boundary bias issues. Most kernel estimators, for instance, have problems with such bounded support because, for points close to the boundaries, they typically place some positive mass outside of the support. \textcolor{blue}{In addition, kernel estimators are not consistent for unbounded densities \cite{geenens2017probit}. This is undesirable since many copula families of interest admit unbounded densities. For example, in the bivariate Gaussian copula case with moderate correlation, the copula density is unbounded in two corners of $[0, 1]^2$. Therefore, any good estimator $\hat{c}$ should be able to cope with such unboundedness. To facilitate this, a variable transformation approach in the kernel estimation of copula densities was proposed \cite{marron1994transformations, geenens2017probit}, which we describe below.}

We define a vector of normal scores $\vect{V} = \left(V_1, \ldots, V_d \right)^\top$ where $V_i = \Phi^{-1}(U_i), 1 \leq i \leq d,$ with $\Phi$ denoting the standard normal CDF and $\Phi^{-1}$ denoting its quantile (probit) function. Given that each $U_i$ is uniformly distributed on $[0, 1]$, we have each $V_i$ distributed as a standard normal variable, although this does not force the joint distribution of $\vect{V}$ to be multivariate normal. Then, the joint density function of $\vect{V}$, given by $g$, may be expressed as follows:
\begin{equation*}
c(u_1, \ldots, u_d) = \frac{g\left\{\Phi^{-1}(u_1), \ldots, \Phi^{-1}(u_d)\right\}}{\phi\left\{\Phi^{-1}(u_1)\right\} \ldots \phi\left\{\Phi^{-1}(u_d)\right\}}, \quad \vect{u} \in [0, 1]^d.
\end{equation*}
The motivation for this probit transformation is as follows: if $c(\vect{u}) > 0$ Lebesgue-a.e. over $[0, 1]^d$, $\vect{V}$ has unconstrained support over $\mathbb{R}^d$ and estimating its density $g$ no longer suffers from boundary issues. In addition, due to its normal margins, one may expect $g$ to be smooth and well-behaved, and its estimation becomes relatively easy and accurate. It is clear that any estimator $\hat{g}$ of $g$ on $\mathbb{R}^d$ will produce a corresponding estimator of the copula density on $(0, 1)^d$:
\begin{equation}
\label{eq:density-probit-estim}
\hat{c}(u_1, \ldots, u_d) = \frac{\hat{g}\left\{\Phi^{-1}(u_1), \ldots, \Phi^{-1}(u_d)\right\}}{\phi\left\{\Phi^{-1}(u_1)\right\} \ldots \phi\left\{\Phi^{-1}(u_d)\right\}}, \quad \vect{u} \in (0, 1)^d.
\end{equation}
Further, when necessary, $\hat{c}$ can also be defined at the boundaries of $[0, 1]^d$ by continuity \cite{geenens2017probit}. This probit transformation trick confers many advantages to the estimator $\hat{c}$, including (i) $\hat{g}$, being an unconstrained density estimator on $\mathbb{R}^d$ does not suffer from boundary bias issues and does not allocate any probability to $\hat{c}$ outside $[0, 1]^d$; (ii) if $\hat{g}$ is a \textit{bona fide} density estimator, i.e., $\hat{g}(\vect{v}) \geq 0$ $\forall \vect{v} \in \mathbb{R}^d$ and $\int_{\mathbb{R}^d} \hat{g}(\vect{v}) d\vect{v} = 1$, then by change of variable $U_i = \Phi(V_i)$ for $1 \leq i \leq d$ we have $\hat{c}(\vect{u}) \geq 0$ $\forall \vect{u} \in [0, 1]^d$ and $\int_{{[0, 1]}^d} \hat{c}(\vect{u}) d\vect{u} = 1$; (iii) if $\hat{g}$ is a uniformly (weak or strong) consistent estimator for $g$, we note that the corresponding $\hat{c}$ inherits the same behavior on any compact subset of $[0, 1]^d$.

The key challenge now lies in obtaining a `good' PDF estimator $\hat{g}$.  Kernel density estimators (KDEs) are commonly used for estimating PDFs. A well-known technical challenge in the KDE method is to determine a kind of optimal bandwidth $H$ in addition to a specifically chosen kernel density \cite{Zeng2018}. Albeit a vast literature exists on this issue of bandwidth tuning, selecting optimal $H$ remains case-dependent and computationally burdensome,  and oftentimes this task involves  a manual user intervention \cite{Silverman1986}. A review of automatic selection methods \cite{Heidenreich2013} recommends a variety of different approaches that are dependent on data set characteristics (including sample size, smoothness, and skewness) and thus are hard to implement properly in practice. In Section \ref{sec:sc_estimation}, we consider an alternative approach to density estimation that relies on fast Fourier transforms.

In summary, using the probit-transformation trick and an FFT-based density estimation method, we are able to obtain the estimated copula density functions $\hat{c}_{\vect{U_X}\vect{U_Y}}$, $\hat{c}_{\vect{U_X}}$, and $\hat{c}_{\vect{U_Y}}$ via Equation \ref{eq:density-probit-estim}. These estimated densities are used to compute our estimator \texttt{fastMI} for $MI$ as described in  Section \ref{sec:np_fastmi}.

\subsection{\emph{\texttt{fastMI}}: fast nonparametric estimation of MI}\label{sec:np_fastmi}
Let $\vect{Z}_i = \left(\vect{X}_i ^\top, \vect{Y}_i ^\top \right)^\top$, with  $\vect{X}_i = \left(X_{1i}, \ldots, X_{pi} \right)^\top$ and $\vect{Y}_i = \left(Y_{1i}, \ldots, Y_{qi} \right)^\top$ for $i \leq i \leq n$ be a random sample drawn from a $d$-variate distribution $f_{\vect{XY}}$. As a preliminary processing step, we define the vectors of empirical probability integral transforms $\vect{U}_{\vect{X}i} = \left(\hat{F}_{\vect{X}_1}(X_{1i}), \ldots, \hat{F}_{\vect{X}_p}(X_{pi}) \right)^\top$ and $\vect{U}_{\vect{Y}i} = \left(\hat{F}_{\vect{Y}_1}(Y_{1i}), \ldots, \hat{F}_{\vect{Y}_q}(Y_{qi}) \right)^\top$. Here, $\hat{F}_{X_i}$ and $\hat{F}_{Y_j}$ are the empirical CDFs of $X_i$ and $Y_j$ respectively for $1 \leq i \leq p$ and $1 \leq j \leq q$. Next, we invoke the probit transformation as described in Section \ref{sec:np_cop} to obtain the estimated copula density functions: $\hat{c}_{\vect{U_X}\vect{U_Y}}$, $\hat{c}_{\vect{U_X}}$, and $\hat{c}_{\vect{U_Y}}$, as described by Equation \ref{eq:density-probit-estim}. These estimated densities are used to  compute our plug-in estimator \texttt{fastMI} estimator using Equation \ref{eq:copula-entropy}. Consequently, \texttt{fastMI} is given by 
\begin{equation}
\label{eq:fastMI}
    \widehat{MI}_{fast} = n^{-1} \sum_{i = 1}^n \ln \left\{ \frac{ \hat{c}_{\vect{U_X}\vect{U_Y}}(\vect{U}_{\vect{X}i}, \vect{U}_{\vect{Y}i})}{ \hat{c}_{\vect{U_X}}(\vect{U}_{\vect{X}i}) \hat{c}_{\vect{U_Y}}(\vect{U}_{\vect{Y}i})} \right\}. 
\end{equation} 
Note that we use \texttt{fastMI} and $\widehat{MI}_{fast}$ interchangeably in this article. In Section \ref{sec:sc_estimation} we discuss a data-driven fast Fourier transform-based density estimation technique that yields consistent and fast estimates of the underlying density function without being encumbered by bandwidth selection issues.

\section{Self-consistent density estimation via fast Fourier transforms} \label{sec:sc_estimation}
A density estimation technique that has no need for user-selected parameter tuning was introduced by \cite{Bernacchia2011} for univariate continuous random variables and later extended to higher dimensions by \cite{OBrien2016}. The density estimator called the self-consistent (SC) estimator, was shown to converge almost surely to the true underlying density for the univariate case. In this section, we present the derivation of the SC estimator for the multivariate case and present proof of almost sure consistency of the SC estimator to the true $d-$dimensional density under mild assumptions. To our knowledge, while proof of consistency for the SC estimator exists for the univariate case, this is the first attempt at proving the consistency of the estimator for the multivariate case. 

The SC estimator has desirable large-sample properties in addition to enjoying greatly improved computations speeds. Using the SC estimator and Equation \ref{eq:density-probit-estim}, we obtain the estimated copula density functions $\hat{c}_{\vect{U_X}\vect{U_Y}}$, $\hat{c}_{\vect{U_X}}$, and $\hat{c}_{\vect{U_Y}}$, which are then used to compute \texttt{fastMI} using Equation \ref{eq:fastMI}.

\subsection{Formulation}
Let $\mathcal{S} = \left\{ \vect{Z}_j \in \mathcal{Z} \subseteq \mathbb{R}^d, j = 1,\ldots,n\right\}$ denote a random sample drawn from from PDF ${f}_{\vect{Z}}$. Assume that the true PDF ${f}_{\vect{Z}}$ belongs to the Hilbert space of square-integrable functions, namely $\mathcal{L}^2 = \left\{g: \int g^2(\vect{z})d\vect{z} < \infty \right\}$. We propose to estimate $\hat{f}_{\vect{Z}}$ by the following convolution of an arbitrary kernel $K$ and the set of delta functions centered on the dataset: 
\begin{equation}
\begin{aligned}
 \label{eqn:kde}
     \hat{f}_{\vect{Z}}(\vect{z}) &:= {n}^{-1} \sum_{j = 1}^{n} K(\vect{z} - \vect{Z}_j) \quad \vect{z} \in \mathbb{R}^d\\ &=  {n}^{-1} \sum_{j = 1}^{n} \int_{\mathbb{R}^d} K(\vect{t}) \delta(\vect{z} - \vect{Z}_j - \vect{t})d\vect{t}\\
     &=  {n}^{-1} \sum_{j = 1}^{n}  \left( K \ast \delta \right) (\vect{z} - \vect{Z}_j),
\end{aligned} 
\end{equation} where $\delta(\vect{z})$ is the Dirac delta function  \cite{Kreyszig_Erwin2020-07-21} and $f \ast g$ denotes the convolution of two functions $f$  and $g$: $(f \ast g)(\vect{t}):=\int_{\mathbb{R}^d} f(\vect{\tau}) g(\vect{t}-\vect{\tau}) d\vect{\tau}.$ Here $K$ is chosen such that the resulting estimator $\hat{f}_{\vect{Z}} \in  \mathcal{L}^2$.

Denote the space of candidate kernel functions by 
\begin{equation*}
    \mathcal{K} := \left\{K: K(\vect{z}) \geq 0; K(\vect{z}) = K(-\vect{z}); \int K(\vect{z})d\vect{z} = 1 \right\}.
\end{equation*}
Given a random sample $\mathcal{S}$, our aim is to identify an optimal kernel $\hat{K} \in \mathcal{K}$, that achieves the minimal mean integrated square error (MISE) between the true density ${f}_{\vect{Z}}$ and an estimator $\hat{f}_{\vect{Z}}$:

\begin{equation}
\begin{aligned}
 \label{eqn:mise}
     \hat{K} &:= \underset{{K} \in \mathcal{K}}{\text{argmin}} \ \text{MISE}(\hat{f}_{\vect{Z}}, {f}_{\vect{Z}}) \\ &= \underset{{K} \in \mathcal{K}}{\text{argmin}} \  \mathbb{E}\left[\int_{\mathbb{R}^d}\{\hat{f}_{\vect{Z}}(\vect{z})-{f}_{\vect{Z}}(\vect{z})\}^{2} d\vect{z}\right],
\end{aligned}
\end{equation} 
where the $\mathbb{E}$ operator denotes expectation taken over the entire support of ${f}_{\vect{Z}}$. 

To perform minimization in Equation \ref{eqn:mise}, we follow a procedure for signal deconvolution \citep{wiener1949extrapolation} via Fourier transforms. First, consider the Fourier transform of the true density $f_{\vect{Z}}$, namely the characteristic function (CF) given by
\begin{equation*}
    \phi(\vect{t}) := \int_{\mathbb{R}^d} f_{\vect{Z}}(\vect{z})\exp(i \vect{t}^\top \vect{z}) d\vect{z}, \quad \vect{t} \in \mathbb{R}^d.
\end{equation*} Then, using the Fourier convolution theorem we obtain
\begin{equation}
    \begin{aligned}
    \label{eqn:fourier_conv}
        \mathcal{F}\{K(\vect{z}-\vect{Z}_j)\}(\vect{t}) &= \mathcal{F}\{K(\vect{z}) \ast \delta(\vect{z}-\vect{Z}_j)\}(\vect{t}) \\ &=\kappa(\vect{t}) \exp(i \vect{t}^\top \vect{Z}_j), \quad \vect{t} \in \mathbb{R}^d,
    \end{aligned}
\end{equation} 
where $\mathcal{F}$ is the Fourier transform operator and $\kappa(\vect{t}):= \mathcal{F}\left\{K(\vect{z})\right\}(\vect{t})$ denotes the Fourier transform of the kernel $K$. From Equation \ref{eqn:fourier_conv} and the linearity of Fourier transforms in Equation \ref{eqn:kde}, we obtain $\hat{\phi}(\vect{t})$, the Fourier transform of the estimate $\hat{f}_{\vect{Z}}$, as follows:
\begin{equation}
    \begin{aligned}
    \label{eqn:phi-hat}
        \hat{\phi}(\vect{t}) &= \mathcal{F}\left\{\hat{f}_{\vect{z}}(\vect{z})\right\}(\vect{t}) \\
        &= \kappa(\vect{t}) \ n^{-1} \sum_{j = 1}^n \exp(i \vect{t}^\top \vect{Z}_j) \\
        &= \kappa(\vect{t}) \mathcal{C}(\vect{t})
    \end{aligned}, \quad \vect{t} \in \mathbb{R}^d,
\end{equation} where $\mathcal{C}(\vect{t}) =  n^{-1} \sum_{j = 1}^n \exp(i \vect{t}^\top \vect{Z}_j)$ is the empirical characteristic function (ECF). Since the data $\left\{ \vect{Z}_1, \ldots, \vect{Z}_n \right\}$ are i.i.d., it is easy to see that the ECF is an unbiased estimator of the corresponding CF, i.e., $\mathbb{E}\left\{\mathcal{C}(\vect{t}) \right\} = {\phi}(\vect{t})$. It can also be shown that $\mathbb{E} \ \lvert \mathcal{C}(\vect{t}) -  {\phi}(\vect{t}) \rvert^2  = \left(1 - \lvert {\phi}(\vect{t}) \rvert^2 \right)/n$. We refer the reader to \cite[Chapter~3]{Ushakov1999} for more details on the ECF and its properties. It follows that $\mathbb{E}\left\{\hat{\phi}(\vect{t}) \right\} = \kappa(\vect{t}){\phi}(\vect{t})$ for $\vect{t} \in \mathbb{R}^d$.

The MISE in Equation \ref{eqn:mise} corresponds to the mean-square distance between the true density $f_{\vect{Z}}$ and the estimate $\hat{f}_{\vect{Z}}$, in terms of the Euclidean metric in the Hilbert space $\mathcal{L}^2$. We rewrite Equation \ref{eqn:mise} in Fourier space using Parseval's theorem as follows:
\begin{equation}
\begin{aligned}
 \label{eqn:mise_fourier}
     \text{MISE}(\hat{f}_{\vect{Z}}, {f}_{\vect{Z}}) = (2\pi)^{-d} \ \mathbb{E}\left[\int_{\mathbb{R}^d}\lvert \hat{\phi}(\vect{t})-{\phi}(\vect{t})\rvert^{2} d\vect{t}\right].
\end{aligned}
\end{equation} Since ${f}_{\vect{Z}}(\vect{z}), \hat{f}_{\vect{Z}}(\vect{z}), {\phi}(\vect{t}), \hat{\phi}(\vect{t}) \in \mathcal{L}^2$, we may interchange the expectation and integral operations \cite{billingsley} and rewrite the MISE in Equation \ref{eqn:mise_fourier} as follows:
\begin{equation}
\begin{aligned}
 \label{eqn:mise_fourier_quadratic}
     \text{MISE}(\hat{f}_{\vect{Z}}, {f}_{\vect{Z}})
     &= (2\pi)^{-d} \ \int_{\mathbb{R}^d}\left[\mathbb{E} \lvert \hat{\phi}(\vect{t})-{\phi}(\vect{t})\rvert^{2} d\vect{t}\right] \\
     &= (2\pi)^{-d} \ \int_{\mathbb{R}^d}\left[\mathbb{E} \lvert \hat{\phi}(\vect{t}) - \mathbb{E}\left\{ \hat{\phi}(\vect{t})\right\} + \mathbb{E}\left\{ \hat{\phi}(\vect{t})\right\}  -{\phi}(\vect{t})\rvert^{2} d\vect{t}\right] \\
     &= (2\pi)^{-d} \int_{\mathbb{R}^d}  \left[n^{-1} \lvert \kappa (\vect{t}) \rvert^2 \left\{1 -\lvert{\phi (\vect{t})}\rvert^2 \right\} +  \lvert{\phi (\vect{t})}\rvert^2 \lvert 1 - \kappa (\vect{t}) \rvert^2 \right]  d\vect{t}.
\end{aligned}
\end{equation} 
Since the integrand in the last line of Equation \ref{eqn:mise_fourier_quadratic} is quadratic in $\kappa$ it is straightforward to find the optimal Fourier-transformed kernel ${\kappa}_{OPT}$ that minimizes MISE, by equating the functional derivative of MISE with respect to $\kappa$ to zero and solving for ${\kappa}_{OPT}$, given by
\begin{equation}
\begin{aligned}
 \label{eqn:kappa_optim_ap}
     {\kappa}_{OPT}(\vect{t}) = \frac{n}{n -1 + \lvert \phi(\vect{t}) \rvert^{-2}}, \quad \vect{t} \in \mathbb{R}^d.
\end{aligned}
\end{equation} Equation \ref{eqn:kappa_optim_ap} reveals that a unique `optimal' Fourier transformed kernel can be derived as a function of the power spectrum of the (unknown) density that is to be estimated. Although this finding was first reported in \cite{Watson1963}, the result alone is unfortunately of little use since the power spectrum $\phi$ of the true density $f_{\vect{Z}}$ is not known. We follow the suggestions in \cite{Bernacchia2011} and take a step further to plug in to Equation \ref{eqn:phi-hat}, the transformed kernel $\kappa_{OPT}$ obtained in Equation \ref{eqn:kappa_optim_ap} to write the Fourier transform $\hat{\phi}$ of the density estimate $\hat{f}_{\vect{z}}$ as follows
\begin{equation}
\begin{aligned}
 \label{eqn:phi_optim_ap}
     \hat{\phi}(\vect{t}) =  \mathcal{C}(\vect{t}) {\kappa}_{OPT}(\vect{t})  =  \frac{n \mathcal{C}(\vect{t})}{n - 1 + \lvert \phi(\vect{t}) \rvert^{-2}}, \quad \vect{t} \in \mathbb{R}^d.
\end{aligned}
\end{equation} Then, an iterative procedure proposed by \cite{Bernacchia2011, OBrien2016} may be applied to determine the exact fixed point of Equation \ref{eqn:phi_optim_ap}, with the details described in Section \ref{sec:density_sce}.

\subsection{Multivariate self-consistent estimation} \label{sec:density_sce}
An iterative procedure is defined by the following sequence of estimates, begun with an initial guess $\tilde{\phi}_0(\vect{t})$: 
\begin{equation}
\begin{aligned}
 \label{eqn:phi_optim_iterative}
     \tilde{\phi}_{n+1}(\vect{t}) = \frac{n \mathcal{C}(\vect{t})}{n - 1 + \lvert \tilde{\phi}_{n}(\vect{t}) \rvert^{-2}}, \quad \vect{t} \in \mathbb{R}^d.
\end{aligned}
\end{equation} The convergent point from Equation \ref{eqn:phi_optim_iterative} yields an estimate $\tilde{\phi}_{SC}(\vect{t})$ which satisfies
\begin{equation}
\begin{aligned}
     \label{eqn:phi_optim_sc}
     \tilde{\phi}_{SC}(\vect{t}) = \frac{n \mathcal{C}(\vect{t})}{n - 1 + \lvert \tilde{\phi}_{SC}(\vect{t}) \rvert^{-2}}, \quad \vect{t} \in \mathbb{R}^d.
\end{aligned}
\end{equation} The existence of $\tilde{\phi}_{SC}(\vect{t})$ is warranted as Equation \ref{eqn:phi_optim_sc} has two fixed and unique non-null solutions. We show that Equation \ref{eqn:phi_optim_sc} has only one stable non-null solution. Noting that $\tilde{\phi}_{SC}$ is complex-valued, we take the absolute value of Equation \ref{eqn:phi_optim_sc}. When the null solution $\tilde{\phi}_{SC} = 0$ is removed, we obtain a simple quadratic equation: 
\begin{equation*}
\begin{aligned}
     \left( n - 1\right) \lvert \tilde{\phi}_{SC}(\vect{t}) \rvert^2 = n \lvert \mathcal{C}(\vect{t}) \rvert \lvert \tilde{\phi}_{SC}(\vect{t}) \rvert.
\end{aligned}
\end{equation*} Provided $\lvert \mathcal{C}(\vect{t}) \rvert \geq 4(n - 1)/n^2$, the above equation has two solutions denoted by the superscripted $\lvert \tilde{\phi}_{SC}(\vect{t})^{\pm} \rvert$,
\begin{equation*}
\begin{aligned}
     \lvert\tilde{\phi}_{SC}(\vect{t})^{\pm}\rvert=\frac{n \lvert \mathcal{C}(\vect{t})\rvert }{2(n-1)}\left[1 \pm \sqrt{\left\{1-\frac{4(n-1)}{n^2|\mathcal{C}(\vect{t})|^2}\right\}}\right].
\end{aligned}
\end{equation*} When substituted in Equation \ref{eqn:phi_optim_sc}, they return the solution for $\tilde{\phi}_{SC}(\vect{t})^{\pm}$ of the form:
\begin{equation*}
\begin{aligned}
    \tilde{\phi}_{SC}(\vect{t})^{\pm} =\frac{n  \mathcal{C}(\vect{t}) }{2(n-1)}\left[1 \pm \sqrt{\left\{1-\frac{4(n-1)}{n^2|\mathcal{C}(\vect{t})|^2}\right\}}\right], \quad \vect{t} \in \mathbb{R}^d \cap \left\{\vect{t}: \lvert \mathcal{C}(\vect{t}) \rvert \geq 4(n - 1)/n^2\right\}.
\end{aligned}
\end{equation*} Since these two solutions only depend on the ECF $\mathcal{C}$, they may be used to estimate $\phi$. Below, we identify a stable solution to be used in Equation \ref{eqn:phi_optim_ap}.

Whereas $\tilde{\phi}_{SC}^{+}$ is normalized, $\tilde{\phi}_{SC}^{-}$ is not; that is, $\mathcal{C}(\vect{0}) = 1$ implies $\tilde{\phi}_{SC}^{+}(\vect{0}) = 1$, which is desirable; in contrast $\tilde{\phi}_{SC}^{-}(\vect{0}) = 1/(n-1)$, which is undesirable. Furthermore, we compute the derivative 
\begin{equation*}
    \begin{aligned}
        \left.\frac{\mathrm{d}\lvert\tilde{\phi}_{n+1}\rvert}{\mathrm{d}\lvert\tilde{\phi}_n\rvert}\right|_{\lvert\tilde{\phi}_n\rvert=\lvert\tilde{\phi}_{SC}^{ \pm}\rvert}=1 \mp \sqrt{\left\{1-\frac{4(n-1)}{n^2|\mathcal{C}(t)|^2}\right\}}. 
    \end{aligned}
\end{equation*} 
Thus, under $\lvert \mathcal{C}(\vect{t}) \rvert \geq 4(n - 1)/n^2$, $\tilde{\phi}_{SC}^{+}$ has a derivative smaller than one, indicating a stable equilibrium point, whereas $\tilde{\phi}_{SC}^{-}$ has a derivative larger than one, indicating instability. In summary, we choose $\hat{\phi}_{SC} = \tilde{\phi}_{SC}^{+}$ as the solution to Equation \ref{eqn:phi_optim_ap} in the construction of the estimate. That is,  
\begin{equation}
\begin{aligned}
\label{eqn:phi_optim_ap_solved}
    \hat{\phi}_{SC}(\vect{t}) =\frac{n  \mathcal{C}(\vect{t}) }{2(n-1)}\left[1 - \sqrt{\left\{1-\frac{4(n-1)}{n^2|\mathcal{C}(\vect{t})|^2}\right\}}\right] \mathbb{I}_{A_n}(\vect{t}), \quad \vect{t} \in \mathbb{R}^d.
\end{aligned}
\end{equation} where $A_{n}$ serves as a low-pass filter that ensures the stability of the estimation process. Since $\hat{\phi}_{SC}$ and $A_n$ are bounded (see remarks below), Equation \ref{eqn:phi_optim_ap_solved} can be antitransformed back to the estimate in real space, given as follows:
\begin{equation}
\begin{aligned}
\label{eqn:f_optim_ap_solved}
    \hat{f}_{SC}(\vect{x})= (2\pi)^{-d} \int_{\mathbb{R}^d} \hat{\phi}_{SC}(\vect{t}) \exp (-i \vect{t}^\top \vect{x}) d\vect{t}. 
\end{aligned}
\end{equation}

\emph{Remark 1: } The purpose of the filter $\mathbb{I}_{A_{n}}(\vect{t})$ is to define a Fourier-based low-pass filter on the ECF $\mathcal{C}(\vect{t})$ that yields a stable optimal estimate in the minimum MISE sense. Primarily, the set $A_{n}$ is specified such that: 

\begin{equation}
\label{eq:filter}
A_{n} = \left\{\vect{t} \in \mathbb{R}^d: \left|\mathcal{C}(\vect{t}) \right|^2 \geq \mathcal{C}_{\text{min}}^2 = \frac{4({n}-1)}{{n}^2} \right\}.
\end{equation} 
Here, the primary filter is necessary for the stability of the iteration method, since the lower bound $C_{\text{min}}$ can ensure a well-defined square root term in Equation \ref{eqn:phi_optim_ap_solved}. Moreover, according to \cite{Bernacchia2011}, the set $A_{n}$ may exclude an additional small subset of frequencies to produce a smoother density estimate $\hat{f}_{\vect{Z}}$. In order for $\hat{f}_{\vect{Z}}$ to converge to the true density ${f}_{\vect{Z}}$ as ${n}$ increases, we require that this set of additionally excluded frequencies must shrink, so that the set $A_{n}$ of included frequencies  grows with increasing ${n}$. 

\emph{Remark 2: } According to \cite{OBrien2016}, the multivariate ECF $\mathcal{C}(\vect{t})$ consists of a finite set of contiguous hypervolumes denoted by $\left\{HV_1^{n}, \ldots, HV_{k_n}^{n}  \right\}$ where $k_{n}$ is a finite integer. Each hypervolume permits `above-threshold' frequency values $\vect{t}$ for which the constraint in Equation \ref{eq:filter} holds. Note that at least one such contiguous hypervolume containing $\vect{t} = \vect{0}$ is guaranteed to exist since $\mathcal{C}(\vect{0}) = 1$ due to normalisation and the primary filter $A_{n}$ has a lower bound $\mathcal{C}_{\text{min}} \leq 1$. Following the suggestion by \cite{OBrien2016} we employ the \emph{lowest contiguous hypervolume} filter, choosing the only hypervolume centered at $\vect{t} = \vect{0}$, which we denote as $HV_1^{n}$ for notational convenience. We make the following observations about $HV_1^{n}$:
\begin{enumerate}
    \item The set of frequencies included in the lowest contiguous hypervolume filter is bounded above since the frequencies will always be contained within a finite-sized hypervolume around the origin.
    \item The volume of the lowest contiguous hypervolume filter  grows as the number of data points ${n}$ increases, implying more frequencies are included for larger sample sizes.
\end{enumerate}
The resulting filter satisfies the convergence conditions described by \cite{Bernacchia2011}. Hence, we set $A_{n} = HV_1^{n}$, and study convergence of $\hat{f}_{SC}$ to the true ${f}_{\vect{z}}$ as ${n}$ increases. For notational convenience, let $\bar{A}_{n}$ denote the complement set of $A_{n}$ and $\mathcal{V}(A_{n})$ denote the volume of $A_{n}$.

When compared to the classical kernel estimation approach \citep{Silverman1986} that assumes a specific form of the kernel with a need of tuning bandwidth, the advantage of the above SCE lies in making minimal assumptions on the functional form of $K$, while determining the estimate $\hat{\phi}_{SC}$ as a function of the ECF $\mathcal{C}$, along with a well-specified low-pass frequency filter $A_n$ entirely by a data-driven approach. The proposed optimization via Equation \ref{eqn:phi_optim_ap} enjoys computational efficiency when `finding' the functional form of the optimal kernel.

\section{Theoretical guarantees} 
\label{sec:theory}

\subsection{Large sample properties}
We now present key large-sample properties for the SCE estimator and the subsequent plug-in estimator of $MI$,  beginning with Theorem \ref{thm:strong_conv1} that establishes strong consistency of the SCE estimator $\hat{f}_{SC}$ at all points on the support of ${f}_{\vect{Z}}$.

\begin{theorem}
\label{thm:strong_conv1} Let the true density ${f}_{\vect{Z}}$ be square integrable and its corresponding Fourier transform $\phi$ be integrable, then the self consistent estimator $\hat{f}_{SC}$, which is defined by  Equation \ref{eqn:f_optim_ap_solved}, converges almost surely to ${f}_{\vect{Z}}$ as $n \rightarrow \infty$, under the additional  assumptions $\mathcal{V}(A_{n}) \rightarrow \infty$, $\mathcal{V}(A_{n})/\sqrt{{n}} \rightarrow 0$ and $\mathcal{V}(\bar{A}_{n}) \rightarrow 0$ as $n \rightarrow \infty$. Further, assuming ${f}_{\vect{Z}}$ to be continuous on support $\mathbb{R}^d$, we have uniform almost sure convergence of $\hat{f}_{SC}$ to ${f}_{\vect{Z}}$ as $n \rightarrow \infty$. 
\end{theorem}
The proof of Theorem \ref{thm:strong_conv1} is given in Section \ref{subsubsec:proof_thm1}. Note that $\hat{f}_{SC}$  can be used in conjunction with the probit transformation trick described in Section \ref{sec:np_cop} in order to obtain uniformly strong consistent estimators for the copula density functions $\hat{c}_{\vect{U_X}\vect{U_Y}}$, $\hat{c}_{\vect{U_X}}$, and $\hat{c}_{\vect{U_Y}}$ and subsequently obtain the \texttt{fastMI} estimator described by Equation \ref{eq:fastMI}. \texttt{fastMI} is shown to be consistent, as established by Theorem \ref{thm:mi-consistency2} below. 
\begin{theorem}
\label{thm:mi-consistency2} Let the assumptions of Theorem \ref{thm:strong_conv1} hold. Further, we assume the true underlying copula functions ${c}_{\vect{U_X}\vect{U_Y}}$, ${c}_{\vect{U_X}}$, and ${c}_{\vect{U_Y}}$  are bounded away from zero and infinity on their support. Then, the estimator \em{\texttt{fastMI}} given by Equation \ref{eq:fastMI} converges in probability to the true $MI$ given in Equation \ref{eq:copula-entropy} as $n \rightarrow \infty$. 
\end{theorem}
The proof of Theorem \ref{thm:mi-consistency2} is given in Section \ref{subsubsec:proof_thm2}.

\subsection{Test for independence}
On the basis of Theorem \ref{thm:mi-consistency2}, using the estimator $\widehat{MI}_{fast}$, we propose a permutation-based test for independence, i.e., a test for $H_0: MI = 0$ against the alternative $H_a: MI > 0$.

Typically, rejection rules of a test based on large-sample theory require data with large sample sizes, which may not be always available in practice. Leveraging the fast computational speed of our SC estimation method, in this paper we consider a hypothesis testing method that is deemed stable and reliable.  Thus, we implement a permutation-based test as it is known to give a precise finite-sample distribution of the test statistic for even small samples \cite{Manly2018}. For a random sample of $n$ observations, $S = \left\{(\vect{X}_1 , \vect{Y}_1 ), \ldots, (\vect{X}_n , \vect{Y}_n )\right\}$, let $\{\delta(1), ,\ldots, \delta(n)\}$ be a random permutation of $\{1,\ldots, n\}$. Based on $\delta$-permuted data set $S_\delta = \left\{(\vect{X}_1 , \vect{Y}_{\delta(1)}), \ldots, (\vect{X}_n , \vect{Y}_{\delta(n)} )\right\}$, we calculate the corresponding estimate $\widehat{MI}_{fast}^{\delta}$. On repeating the above procedure $r$ times, $\mathcal{T}_{perm} = \left\{\widehat{MI}^{\delta_1}_{fast}, \ldots, \widehat{MI}^{\delta_r}_{fast}\right\}$, a collection of estimates is obtained, which may be used to approximate the null distribution of $\widehat{MI}_{fast}$, under the null hypothesis $H_0: \vect{X}$ and $\vect{Y}$ are independent. At significance level $\alpha$, we reject the null hypothesis when $\widehat{MI}_{fast}$ based on the original data is greater than the $(1-\alpha)$th empirical quantile of $\mathcal{T}_{perm}$. 

\section{Simulation Experiments}\label{simulations}

Previous replicable studies \cite{Zeng2018} have compared the performance of the $JMI$-based test with other popular methods, including dCor, HHG and MIC-based tests, and concluded that $JMI$ appears to be the most stable test for independence. 
Further, as a benchmark we compare the performance of our \texttt{fastMI} estimator with the empirical copula-based estimator of $MI$ (named $ECMI$), in which a naive bandwidth-driven kernel density estimation route is taken when estimating the copula functions described in Equation \ref{eq:copula-entropy}.

In this section, our proposed nonparametric estimator is compared with the $ECMI$ and $JMI$  from three distinct but inter-related perspectives: (i) estimation accuracy, (ii) ability to test for independence, and (iii) computation time. We design and implement exhaustive simulation-based experiments for our study. In brief: relative to the $ECMI$ as well as the $JMI$, the \texttt{fastMI} is the more accurate estimator with reduced estimation error and is able to conduct a hypothesis test for independence with a higher power while exhibiting satisfactory type I error control. Reduced computation times further indicate it is more amenable to applications in large data sets when compared to $ECMI$ and $JMI$.

\subsection{Estimation accuracy of MI}\label{simulations-1}
Through extensive simulation studies, we compare the MSE performance of \texttt{fastMI} with $ECMI$ and $JMI$ for a wide range of association patterns for different sample sizes. We generate a sample of $n$ observations drawn from PDF $f_{XY}$ on $\mathbb{R}^d$, specified by the underlying copula and marginal densities. In the data simulation, choices of marginal distributions have little effect, and for numerical convenience, normal marginals are chosen for data generation in our empirical studies.

\begin{figure}[ht]
\centering
\includegraphics[width = \textwidth]{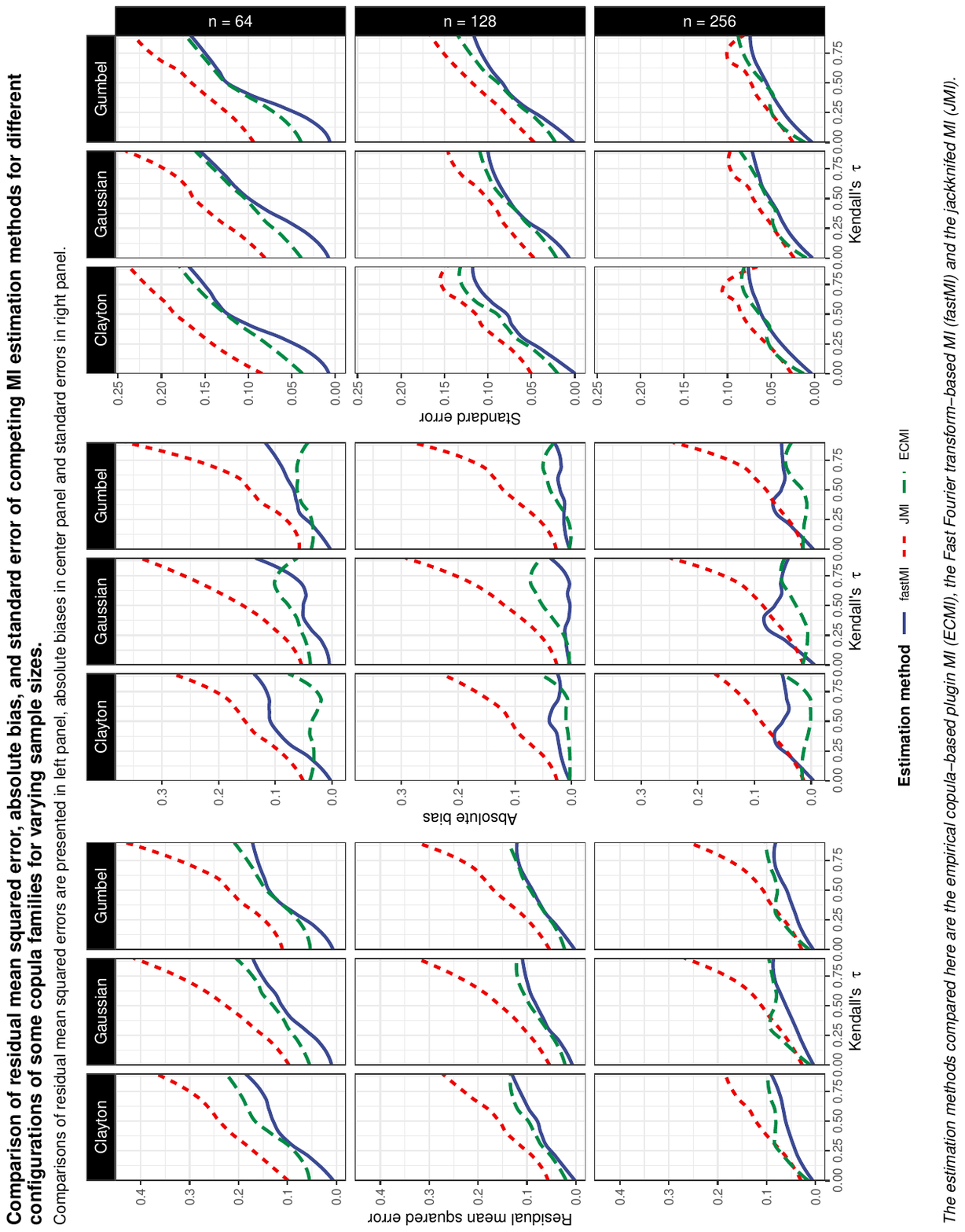}
\caption{}
\label{fig:figure1}
\end{figure}

We first consider the bivariate case ($d = 2$) and restrict ourselves to three separate classes of copula models \cite{joe_2014, czado_2019} -- the symmetric Gaussian copula and two asymmetric Archimedean copula - the Clayton and Gumbel copulas. While the Clayton copula exhibits greater dependence in the negative tail, the Gumbel copula exhibits greater dependence in the positive tail. Each of these copula classes may be specified by fixing the underlying value of Kendall's $\tau$, which in turn, may be used to compute the underlying true $MI$ \cite{Ghalibaf_2020}. For each of the three copula classes considered, we fix $\tau \in \left\{0, 0.1, 0.2, \ldots, 0.9 \right\}$  and generate $n \in \left\{64, 128, 256 \right\}$ bivariate samples. The simulated data are used to compute $ECMI$, $JMI$, and \texttt{fastMI}. The MSEs of all estimators for different models and sample sizes are calculated based on $s = 1000$ replications. Our findings are presented in \hyperref[fig:figure1]{\autoref{fig:figure1}}. 

To study the behavior of our estimators more closely, in \hyperref[tab:table-sim1]{\autoref{tab:table-sim1}} we present information on the following percentage decreases: 
\begin{equation*}
\begin{aligned}
    \text{\% decrease in MSE of \texttt{fastMI} relative to ECMI} &= 100 \times \frac{\text{MSE}(ECMI) -\text{MSE}(\texttt{fastMI})}{\text{MSE}(ECMI)} \\
    \text{\% decrease in MSE of \texttt{fastMI} relative to JMI} &= 100 \times \frac{\text{MSE}(JMI) -\text{MSE}(\texttt{fastMI})}{\text{MSE}(JMI)},
\end{aligned}
\end{equation*} with positive values indicating more accurate performance of \texttt{fastMI} over $ECMI$ and $JMI$ respectively. Higher values indicate reduced estimation error.  

From both the figure as well as the table, we note that \texttt{fastMI} has appreciably lower MSE for all models and sample sizes considered, indicating its superior performance over $JMI$. Further, from \hyperref[fig:figure1]{\autoref{fig:figure1}} we note that while the overall standard error for \texttt{fastMI} is lower than $ECMI$ as well as $JMI$ in almost all the cases considered, the absolute bias performance of \texttt{fastMI} and $ECMI$ are comparable while still being lower than that of $JMI$. Overall, there seems to be a bias-variance trade-off between \texttt{fastMI} and $ECMI$, with \texttt{fastMI} yielding better MSE performance. 

\begin{table}[htp]
\centering
\begin{tabular}{cccc}
\toprule
\multicolumn{1}{c}{\multirow{2}{*}{$\vect{\tau}$}} & \multicolumn{3}{c}{\textbf{Copula}}                         \\  
\multicolumn{1}{c}{}                   & \multicolumn{1}{c}{\textit{Clayton}} & \multicolumn{1}{c}{\textit{Gaussian}} & \multicolumn{1}{c}{\textit{Gumbel}} \\ \midrule
 0.0 & 97 (98) & 28 (81) & 83 (73) \\ 
  0.1 & 83 (72) & 71 (71) & 46 (72) \\ 
  0.2 & 17 (68) & 12 (75) & 14 (80) \\ 
  0.3 & 17 (90) & 97 (99) & 17 (61) \\ 
  0.4 & 78 (70) & 72 (64) & 64 (76) \\ 
  0.5 & 40 (68) & 25 (68) & 22 (73) \\ 
  0.6 & 27 (79) & 36 (91) & 97 (99) \\ 
  0.7 & 5 (54) & 79 (71) & 67 (68) \\ 
  0.8 & 58 (71) & 45 (79) & 32 (73) \\ 
  0.9 & 46 (81) & 42 (83) & 6 (73) \\ 
\bottomrule
\end{tabular}
\caption{Percentage decrease in mean squared error values of \texttt{fastMI} relative to $ECMI$ (\texttt{fastMI} relative to $JMI$) for sample sizes $n = 256$ for all copula models across various levels of association, controlled by Kendall's $\tau$.}
\label{tab:table-sim1}
\end{table}

Now, for the multivariate case ($d > 2$) where both $\vect{X}$ and $\vect{Y}$ are bivariate random vectors following a joint $d-$variate normal distribution with zero mean and dispersion matrix $\Sigma$ of the form
\begin{equation*}
    \Sigma := \begin{bmatrix}
  \Sigma_{XX} & \Sigma_{XY} \\
  \Sigma^\top_{XY} & \Sigma_{YY}
\end{bmatrix}.
\end{equation*} The reason for choosing the multivariate normal distribution is that it is one of the   few multivariate distributions which give rise to a closed-form expression of $MI(\vect{X}, \vect{Y}) = 0.5 \ln \left( \lvert \Sigma_{XX}\rvert \lvert \Sigma_{YY}\rvert / \lvert \Sigma\rvert \right)$ where $\lvert \Sigma \rvert$ denotes the determinant of $\Sigma$. This enables the comparison of the three competing estimators in a meaningful way. We set all marginal variances to be unity and consider different structures of $\Sigma$ including the (i) first-order auto-regressive (AR-1) structure, with the generic $(i, j)-$th element being $\sigma_{ij} = \rho^{\lvert i - j \rvert}$; (ii) compound symmetry (CS) structure, that has off-diagonal elements are set to \textcolor{blue}{$\rho \in (0, 1)$}; (iii) spatial structure with the generic $(i, j)-$th element being $\sigma_{ij} = \exp(-\lvert i - j \rvert/\rho)$; \textcolor{blue}{(iv) two block-correlation matrices that emulate two hierarchies of correlation: a within-block correlation $\rho_w$ and a between-block correlation $\rho$:
\begin{equation*}
    \Sigma_{block} := 
    \begin{bmatrix}
    1 & \rho_w & \rho & \rho \\
    \rho_w & 1 & \rho & \rho \\
    \rho & \rho & 1 & \rho_w \\
    \rho & \rho & \rho_w & 1
  \end{bmatrix}, \quad \rho_w \in \left\{1/3, 2/3 \right\}.
\end{equation*}  We set $\rho_w = 1/3$ for mild within-block correlation and  $\rho_w = 2/3$ for strong within-block correlation.}

\begin{figure}[ht]
\centering
\includegraphics[width = \textwidth]{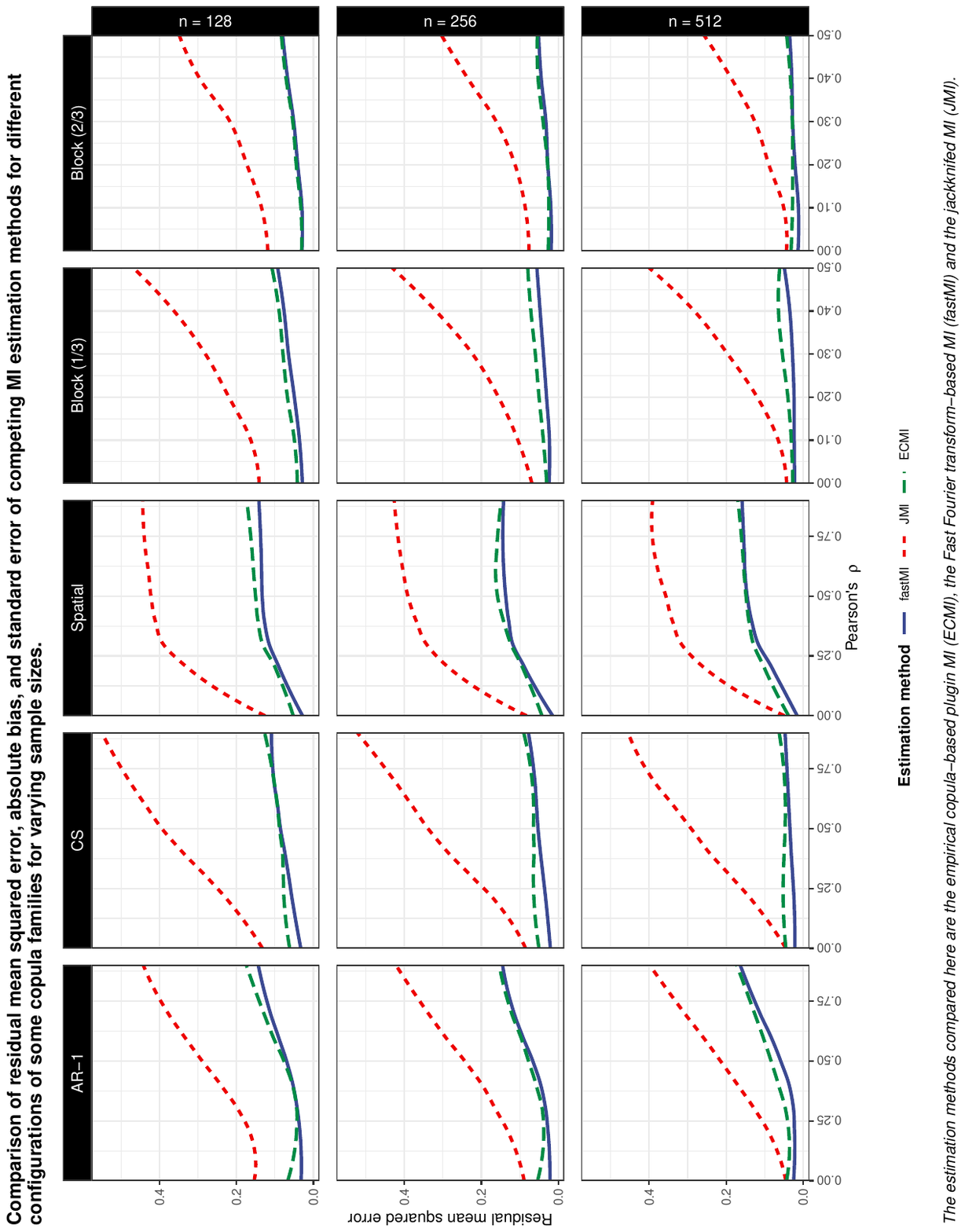}
\caption{}
\label{fig:figure4}
\end{figure}

For each of the five correlation structures described above, we vary $\rho \in  \left\{0, 0.1, 0.2, 0.3, 0.4, 0.5 \right\}$ and compare the accuracy of the competing estimators by means of MSE. Our findings are presented in Figure \ref{fig:figure4}. To study the behavior of our estimators more closely, in \hyperref[tab:table-sim3]{\autoref{tab:table-sim3}} we present information on percentage decrease in MSE of \texttt{fastMI} relative to $ECMI$ and $JMI$ for the different models and with sample size set to $n = 256$. As was the case for lower dimensions, from  \hyperref[fig:figure4]{\autoref{fig:figure4}} and \hyperref[tab:table-sim3]{\autoref{tab:table-sim3}} 
 we note that \texttt{fastMI} has appreciably lower MSE than $ECMI$ and $JMI$ for all models and sample sizes considered, indicating its superior performance.
\begin{table}[]
\centering
\begin{tabular}{cccccc}
\toprule
\multicolumn{1}{c}{\multirow{2}{*}{$\vect{\rho}$}} & \multicolumn{5}{c}{\textbf{Dispersion structure}}                         \\  
\multicolumn{1}{c}{}                   & \multicolumn{1}{c}{\textit{AR-1}} & \multicolumn{1}{c}{\textit{CS}} & \multicolumn{1}{c}{\textit{Spatial}} & \multicolumn{1}{c}{\textit{Block ($\rho_w = 1/3$)}} & \multicolumn{1}{c}{\textit{Block ($\rho_w = 2/3$)}} \\ \midrule
 0.0 & 83 (94) & 80 (95) & 75 (96) & 63 (97) & 47 (97) \\ 
  0.1 & 30 (98) & 84 (95) & 73 (95) & 44 (95) & 32 (96) \\ 
  0.2 & 44 (95) & 24 (92) & 81 (94) & 51 (96) & 7 (90) \\ 
  0.3 & 7 (88) & 19 (87) & 26 (89) & 36 (88) & 65 (95) \\ 
  0.4 & 63 (96) & 58 (97) & 56 (98) & 51 (98) & 50 (94) \\ 
  0.5 & 38 (95) & 4 (94) & 34 (96) & 29 (96) & 13 (97) \\
\bottomrule
\end{tabular}
\caption{Percentage decrease in mean squared error values of \texttt{fastMI} relative to $ECMI$ (\texttt{fastMI} relative to $JMI$) for sample sizes $n = 256$ for different dispersion structures of a 4-dimensional multivariate normal distribution across various strengths of correlation, controlled by Pearson's $\rho$.}
\label{tab:table-sim3}
\end{table}

\subsection{Test for independence} 
\begin{figure}[ht]
\centering
\includegraphics[width = \textwidth]{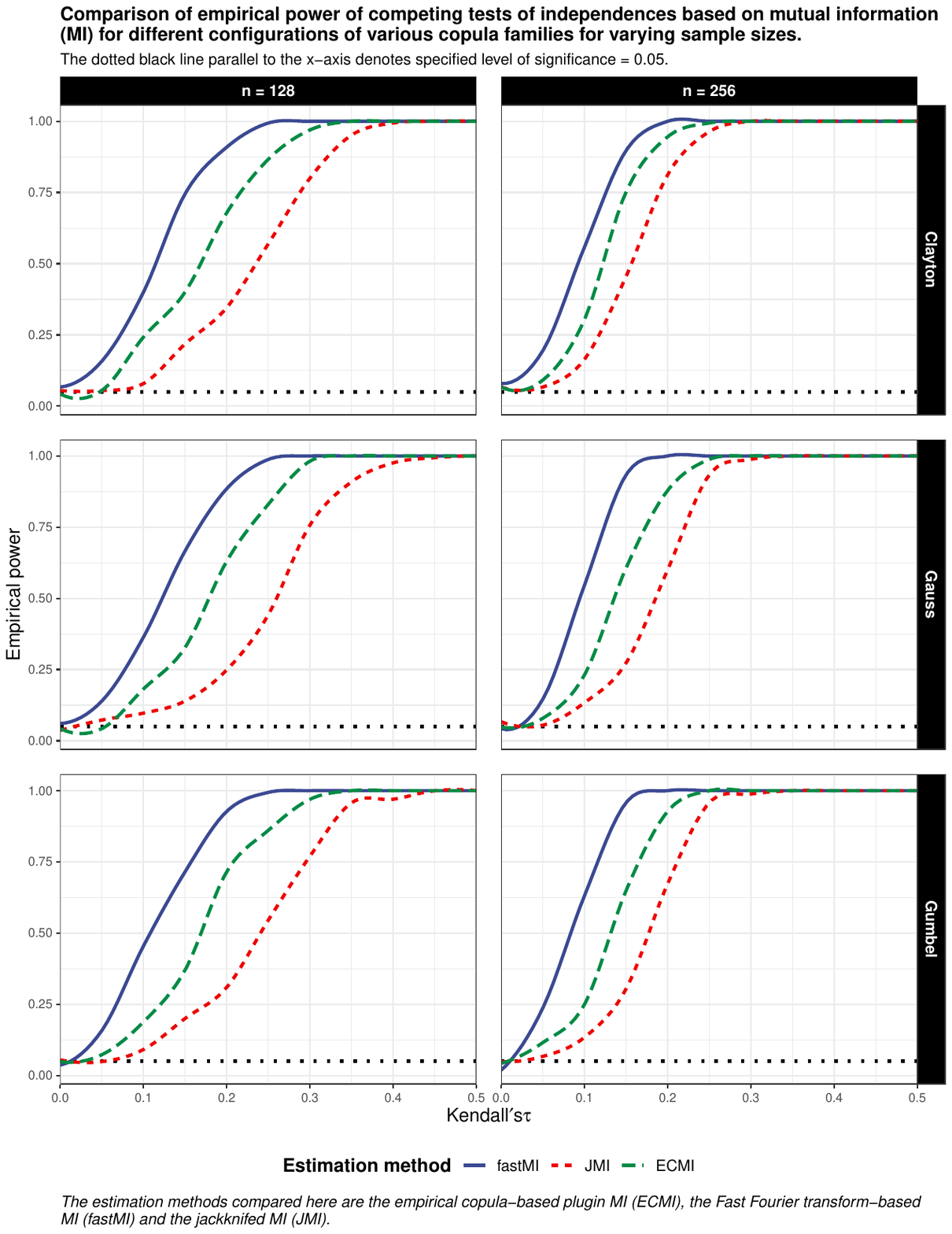}
\caption{}
\label{fig:figure2}
\end{figure}

\begin{figure}[ht]
\centering
\includegraphics[width = \textwidth]{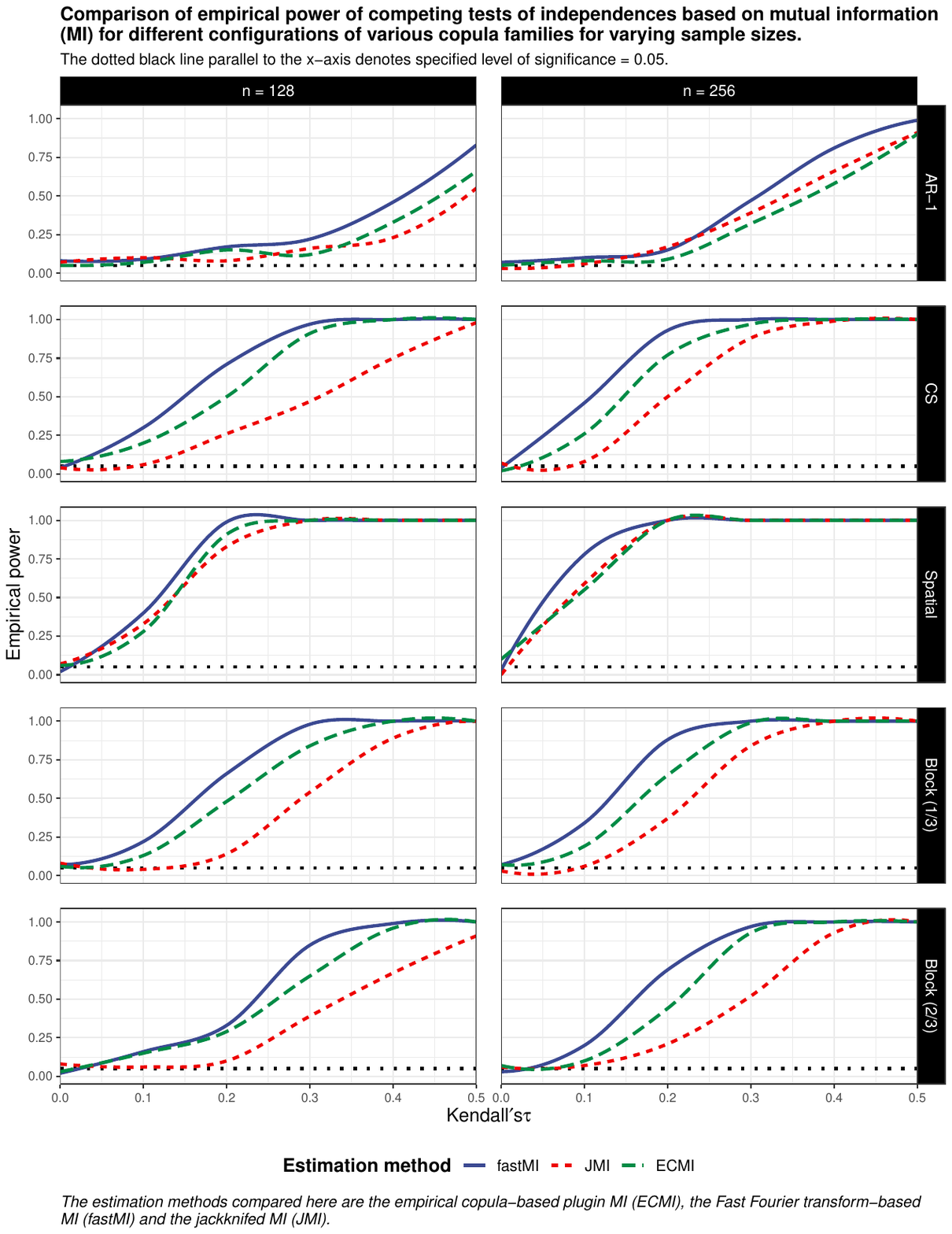}
\caption{}
\label{fig:figure5}
\end{figure}

We compare the permutation-based tests based respectively on our \texttt{fastMI} with $ECMI$ and the $JMI$. For the same bivariate patterns as described in Section \ref{simulations-1}, with $\tau \in \left\{0, 0.05, \ldots, 0.50 \right\}$ ($\tau = 0$ indicates independence), we plot the empirical power curves for the tests at significance level $\alpha = 0.05$ and present our results under different settings in Figure \ref{fig:figure2} for $r = 1000$ permutations. In Table \ref{tab:table-sim2} we present a comparison ($ECMI$ and $JMI$ versus \texttt{fastMI}) of empirical type I error of the permutation-based test for independence in several bivariate distributions for different sample sizes respectively. Note that the type I error rates of all three methods considered are very close to the nominal level $\alpha = 0.05$.
\begin{table}[htp]
\centering
\begin{tabular}{cccc}
\toprule
\multicolumn{1}{c}{\multirow{2}{*}{}} & \multicolumn{3}{c}{\textbf{Copula}}                         \\  
\multicolumn{1}{c}{}                   & \multicolumn{1}{c}{\textit{Clayton}} & \multicolumn{1}{c}{\textit{Gaussian}} & \multicolumn{1}{c}{\textit{Gumbel}} \\ \midrule
 $ECMI$        & $0.04\ (0.07)$    & $0.04\ (0.05)$     & $0.05\ (0.04)$   \\ 
$JMI$    & $0.06\ (0.07)$    & $0.04\ (0.06)$     & $0.06\ (0.07)$   \\ 
\texttt{fastMI} & $0.03\ (0.04)$    & $0.04\ (0.04)$     & $0.03\ (0.04)$   \\\bottomrule
\end{tabular}
\caption{Comparison of empirical type I error of permutation-based test for independence in several bivariate copula families. Using $r = 1000$ permutations, we compare the type I error of $ECMI$, $JMI$, and \texttt{fastMI} for sample sizes $n = 128 \ (n = 256)$.}
\label{tab:table-sim2}
\end{table}
Similarly, for the same multivariate normal patterns, with $\rho \in \left\{0, 0.05, \ldots, 0.50 \right\}$ ($\rho = 0$ indicates independence), we plot the empirical power curves for the tests at significance level $\alpha = 0.05$ and present our results under different settings in Figure \ref{fig:figure5} for $r = 1000$ permutations. In Table \ref{tab:table-sim5} we present a comparison ($ECMI$ and $JMI$ versus \texttt{fastMI}) of empirical type I error of the permutation-based test for independence in several multivariate distributions for different sample sizes respectively. Note that the type I error rates of all three methods considered are very close to the nominal level $\alpha = 0.05$.
\begin{table}[htp]
\centering
\begin{tabular}{cccccc}
\toprule
\multicolumn{1}{c}{\multirow{2}{*}{}} & \multicolumn{5}{c}{\textbf{Dispersion structure}}                         \\  
\multicolumn{1}{c}{}                   & \multicolumn{1}{c}{\textit{AR-1}} & \multicolumn{1}{c}{\textit{CS}} & \multicolumn{1}{c}{\textit{Spatial}} & \multicolumn{1}{c}{\textit{Block ($\rho_w = 1/3$)}} & \multicolumn{1}{c}{\textit{Block ($\rho_w = 2/3$)}} \\ \midrule
  $ECMI$ & 0.05 (0.05) & 0.08 (0.02)  & 0.06 (0.10) & 0.07 (0.06) & 0.03 (0.07) \\
  $JMI$ & 0.07 (0.04) & 0.04 (0.07)  & 0.07 (0.02) & 0.08 (0.08) & 0.08 (0.06) \\
  \texttt{fastMI} & 0.08 (0.07) & 0.04 (0.04)  & 0.04 (0.03) & 0.07 (0.07) & 0.04 (0.03) \\ 
  \bottomrule
\end{tabular}
\caption{Comparison of empirical type I error of permutation-based test for independence in several multivariate normal families. Using $r = 1000$ permutations, we compare the type I error of $ECMI$, $JMI$, and \texttt{fastMI} for sample sizes $n = 128 \ (n = 256)$.}
\label{tab:table-sim5}
\end{table}

\subsection{Computation time}
Since nonparametrically estimating $MI$ is a computationally intensive method, we compare run times of $ECMI$, $JMI$ and \texttt{fastMI}  for various sample sizes for bivariate data. We report the mean and standard deviation of run time (in seconds) in Table \ref{tab:table10}. Of note, for smaller sample sizes, $JMI$ is fastest; in contrast, as sample size increases,  \texttt{fastMI} becomes fast with run times being many fold smaller than both  $ECMI$ and $JMI$. This improvement in run time establishes \texttt{fastMI} as an attractive method for studying association in practice, because it is the case of large sample size that matters in real-world computations.   
\begin{table}[htp]
\centering
\begin{tabular}{cccccc}
\toprule
\multicolumn{1}{c}{\multirow{2}{*}{}} & \multicolumn{5}{c}{\textbf{Sample size ($n$)}}                         \\  
\multicolumn{1}{c}{}                   & \multicolumn{1}{c}{\textit{$250$}} & \multicolumn{1}{c}{$500$} & \multicolumn{1}{c}{$1000$} & \multicolumn{1}{c}{$2500$} & \multicolumn{1}{c}{$5000$} \\ \midrule
$ECMI$    &  $0.695\ (0.141)$  & $1.599\ (0.281)$  & $4.360\ (0.356)$  & $5.368\ (0.308)$ & $6.404\ (0.254)$ \\       
$JMI$    &  $\boldsymbol{0.114\ (0.043)}$  & $\boldsymbol{0.659\ (0.200)}$  & $3.150\ (0.107)$  & $18.446\ (0.116)$ & $62.454\ (4.601)$ \\
\texttt{fastMI} &  $0.355\ (0.088)$ & $0.750\ (0.213)$ & $\boldsymbol{1.199\ (0.135)}$ & $\boldsymbol{2.964\ (0.081)}$  & $\boldsymbol{5.952\ (0.125)}$  \\ \hline
\end{tabular}
\caption{Mean (standard deviation) computation time (in seconds) of $JMI$ and \texttt{fastMI} for bivariate data of varying sample size ($n$) for $s = 100$ iterations.}
\label{tab:table10}
\end{table}

\section{Real Data Analysis}\label{rda}
\begin{figure}[ht]
\centering
\includegraphics[width = \textwidth]{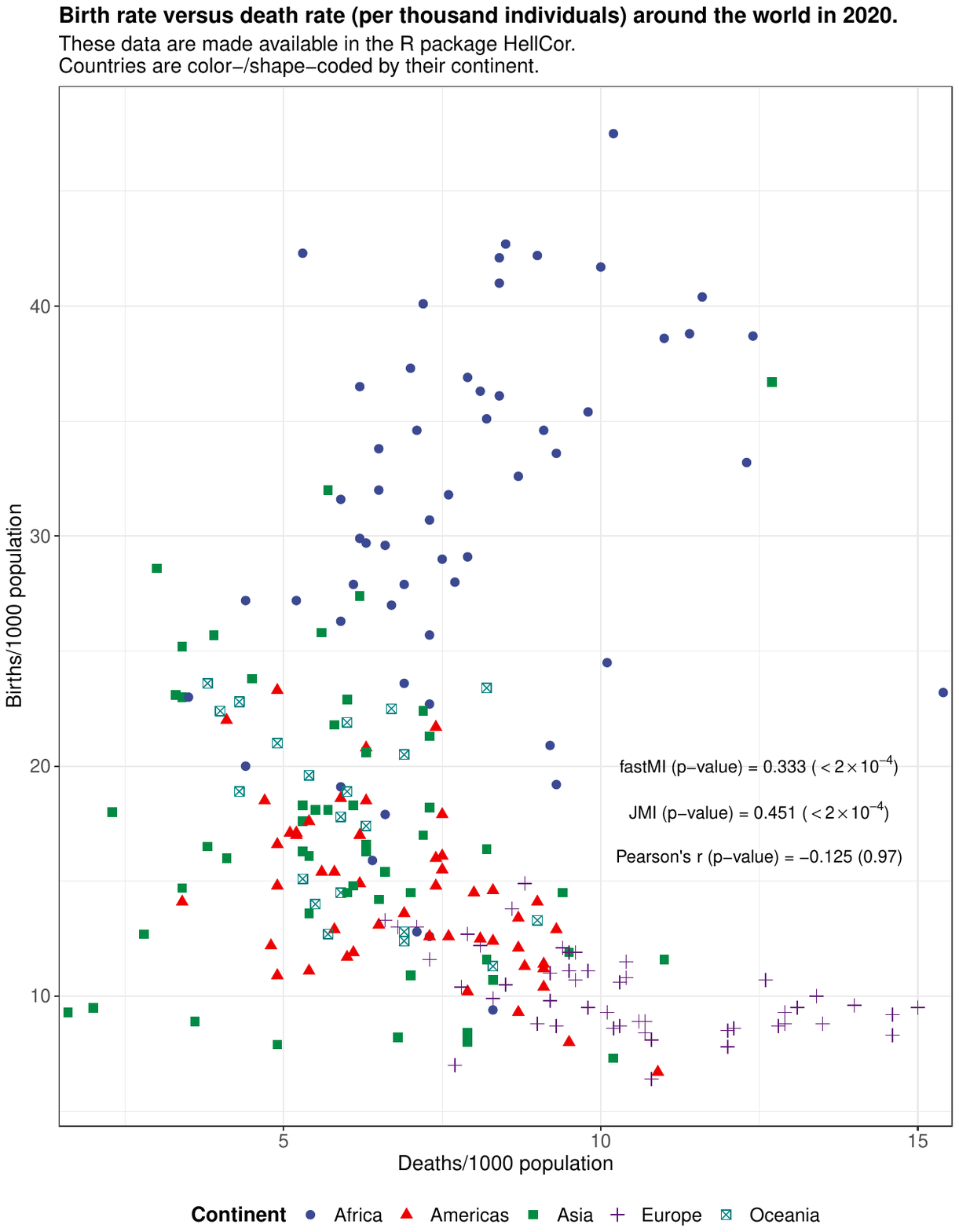}
\caption{}
\label{fig:figure3}
\end{figure}

To illustrate an application of our method in practice, we re-analyse the dependence between $X$ = `death rate' and $Y$ = `birth rate' in 229 countries and territories around the world in the first trimester of 2020 \cite{Geenens_2020}. This dependence was previously found in \cite{Geenens_2020} to be complex with a departure from linearity and monotonicity.  $(X, Y)$ denote the number of deaths and births per year per 1000 individuals in the country. Figure \ref{fig:figure3} presents a scatterplot of the data $(X_i, Y_i)$ denoting the number of deaths and births per year per 1000 individuals in the $i$th country, indicated by one point on the plot. It demonstrates an interesting `C' shape. For ease of exposition, we stratified countries by the continent they belong to, marked by different symbols in the figure. Note that we clubbed North American and South American countries together in Figure \ref{fig:figure3}; this clustering has no role to play in our analysis and is used only for ease of visualization. We may, indeed by simple data visualization, expect a strong association between these two variables whose data points are not randomly distributed but show a clear C-shape. 

A closer investigation by \cite{Geenens_2020} reveals the presence of two possible and opposite relationship patterns. First, it shows an `decreasing' trend from `moderate' birth rate to `low' birth rate as death rate increases, for the industrialized countries (mostly Europe, Oceania and the Americas). Second, an `increasing' trend is present from `moderate' birth rate to `high' birth rate as death rate increases, mostly for African nations. The downward trend is more pronounced, while the increasing trend is more diffused.

The strength of such nonlinear dependence and non-monotonic patterns is hardly captured by Pearson’s correlation whose estimate is $\hat{r} = -0.125$, implying an insignificant association with $p$-value equal to $0.97$ at level $\alpha = 0.05$. In contrast, both \texttt{fastMI} (estimate of $MI$ $ = 0.333$) and $JMI$ (estimated $MI$ $ = 0.451$) report significant association with $r = 5000$ permutations. We find the $p$-values for both statistic to be less than $2 \times 10^{-4}$ via the permutation testing approach at level $\alpha = 0.05$.

We further note the disparity between estimated values of $MI$ using \texttt{fastMI} and $JMI$ estimators, respectively. Based on our knowledge learned from the simulation studies in Section \ref{simulations-1}, namely \texttt{fastMI}-based estimate having reduced estimation error, it is rational to deduce that the $JMI$ method overestimated the strength of association between death rate and birth rate from the data, when compared with our \texttt{fastMI} method. 
This may lead to erroneous inference in other applications.
This example illustrates the advantage of \texttt{fastMI} as being a useful and meaningful measure of dependence  to capture complex nonlinear relationships.

\section{Concluding Remarks}\label{discussion}
This paper develops a fast and consistent $MI$ estimator through a powerful nonparametric copula estimation approach. Through extensive simulation studies, we have demonstrated that the proposed nonparametric method has several desirable properties, outperforming the current gold standard, the $JMI$ as well as the benchmark $ECMI$. Since the proposed method is nonparametric, it does not assume any parametric copula models and works under minimal model assumptions about relationships.  An advantage of the $MI$-based approach is that the quantification of dependence is little dependent on marginal distributions, in which by virtue of its focus on the copula, the proposed method only takes into account the intrinsic association between variables without suffering from potential irregularities in the marginal distributions. An appealing technical advance pertains to that we overcome issues that plague bandwidth-based $MI$ estimators, such as bandwidth selection and slow computation times for large data sets. \texttt{fastMI} relies on a data-adaptive Fast Fourier transformation-based approach to nonparametrically estimating the underlying copula structures. Its run time is many-fold faster than both $ECMI$ as well as $JMI$ for large data sets. \texttt{fastMI} exhibits reduced estimation error and provides a more powerful test of independence than both $ECMI$ and $JMI$. 

While typically used to study pairwise association, $MI$ may be extended in a multivariate setting. However, as evidenced by the discussion below, $MI$ may lose some of its original interpretability, which affects its legitimate use in testing for independence.  Let consider a $d$-dimensional random vector $\vect{X} = \left(X_1, \ldots, X_d \right)^\top$ with joint density function $f_{\vect{X}}$ with marginal density functions $\left\{f_{1},\ldots,f_{d}\right\}$.   It follows from Sklar's theorem that there exists a unique associated copula density given by
\begin{equation*}
c\left(u_1, \ldots, u_d\right)=\frac{f_{\vect{X}}\left(F_{1}^{-1}\left(u_1\right), \ldots, F_d^{-1}\left(u_d\right)\right)}{\prod_{i=1}^d f_i\left(F_i^{-1}\left(u_i\right)\right)}.
\end{equation*}
Here copula density $c(u_1, \ldots, u_d)$ can in turn, be used to define $MI(X_1,\ldots, X_d)$, known as the multivariate mutual information ($MMI$). Other names of $MMI$ include  total correlation \citep{watanabe1960} or multi-information \citep{studenya1998}. Basically, $MMI$ quantifies the amount of information shared among the different random variables, which characterizes the relatedness of the random variables in a group.
\cite{ting1962} proved the possible negativity of $MMI$ when $d \geq 3$. Thus, $MMI$ loses one key property of non-negativity, which makes $MMI$ illegitimate to quantify the size of information (which is believed to be non-negative). $MMI$ is deemed undesirable for its role as a metric of information volume.  

Nevertheless, mathematically, $MMI$ may still be used as a tool to test for independence. From Theorem 2 of \cite{baudot2019}, we know that $d$ variables $X_1,\ldots, X_d$ are mutually independent if and only if all $MMI$s of $2^d - d - 1$ sub-vectors vanish, i.e., $MI(X_{i_1}, \ldots, X_{i_k}) = 0$ with $d \geq k \geq 2$, where the indices $\left\{i_1, \ldots, i_k \right\}$ are a subset of size $k$ of $\left\{1, \ldots, d \right\}$. This would lead to an unduly tedious testing procedure if $MMI$s were used to investigate independence. Some alternative forms of MI are worth future exploration with no sacrifice of non-negativity and avoidance of exhaustive tests.

In conclusion, due to its adaptability to complex associations and robustness to sample size, \texttt{fastMI} may be potentially applicable in a wide range of practical problems where complex non-linearities \cite{Safaai2018} are of particular interest. To encourage use of our proposed toolkit for estimation and testing purposes, we developed an \textsf{R} package \texttt{fastMI} that is publicly available on  \href{https://github.com/soumikp/fastMI}{Github}. 

\section{Technical details}
\label{sec:tech} 

\subsection{Proof of Theorem \ref{thm:strong_conv1}: Large sample behaviour of the optimal density estimator.}
\label{subsubsec:proof_thm1}
In this section we prove Theorem \ref{thm:strong_conv1} which establishes strong consistency of the estimator $\hat{f}_{SC}$ to the true density ${f}_{\vect{Z}}$ as ${n} \rightarrow \infty$ at all points on the support of ${f}_{\vect{Z}}$. 

Note the frequency filter $A_{n}$, its complement $\bar{A}_{n}$ and its volume $\mathcal{V}(A_{n})$ are described in Remarks 1 and 2 in Section \ref{sec:density_sce}. Since the true density ${f}_{\vect{z}}$ and the estimator $\hat{f}_{SC}$ are both square-integrable, we can express them in terms of their corresponding Fourier transforms $\phi$ and $\hat{\phi}$ respectively. Since the characteristic function is integrable, we have, $\int \left| \phi(\vect{t}) \right| d\vect{t} < \infty.$ Through the following sequence of inequalities, we are able to establish an upper bound for the absolute error $\left| \hat{f}_{SC}(\vect{z})-{f}_{\vect{z}}(\vect{z}) \right|$ 
for any $z \in \mathcal{Z}$. By definition, note that $\hat{\phi}(\vect{t}) = 0$ for $\vect{t} \notin A_{n}$. To establish Theorem \ref{thm:strong_conv1}, it is sufficient to show that the upper bound of the absolute error given below tends to zero as ${n} \rightarrow \infty$. We have:
{\allowdisplaybreaks
\begin{align}
    \label{eq:pushright2}
        \left| \hat{f}_{SC}(\vect{z})-{f}_{\vect{z}}(\vect{z}) \right| \notag &= \left| (2 \pi)^{-d}\int_{\mathbb{R}^d} \exp (-\mathrm{i} \vect{t}^\prime \vect{z})\left\{\hat{\phi}(\vect{t})-\phi(\vect{t})\right\} \mathrm{d} \vect{t}\right| \notag\\
        &\leq (2 \pi)^{-d} \int_{\mathbb{R}^d} \lvert \exp (-\mathrm{i} \vect{t}^\prime \vect{z}) \rvert \lvert  \hat{\phi}(\vect{t})-\phi(\vect{t}) \rvert \mathrm{d} \vect{t} \notag \\
        &= (2 \pi)^{-d} \int_{\mathbb{R}^d} \lvert  \hat{\phi}(\vect{t})-\phi(\vect{t}) \rvert \mathrm{d} \vect{t} \notag\\
        &= (2 \pi)^{-d} \int_{A_{n_1}}  \lvert  \hat{\phi}(\vect{t})-\phi(\vect{t}) \rvert \mathrm{d} \vect{t} + 
         (2 \pi)^{-d} \int_{\bar{A}_{n}}  \lvert  \phi(\vect{t}) \rvert \mathrm{d} \vect{t} \notag\\
         &\leq (2 \pi)^{-d} \int_{A_{n}}  \lvert  \hat{\phi}(\vect{t})- \mathcal{C}(\vect{t}) \rvert \mathrm{d} \vect{t} + 
         (2 \pi)^{-d} \int_{A_{n}}  \lvert \mathcal{C}(\vect{t}) -  \phi(\vect{t}) \rvert \mathrm{d} \vect{t} + 
         (2 \pi)^{-d} \int_{\bar{A}_{n}}  \lvert  \phi(\vect{t}) \rvert \mathrm{d} \vect{t} \notag\\
         &:= D_1 + D_2 + D_3.
\end{align}} Under the assumptions, $\lim_{{n} \rightarrow \infty} \ \mathcal{V}\left(A_{n}\right) =  \infty$ and $\lim_{{n} \rightarrow \infty} \ \mathcal{V}\left(\bar{A}_{n}\right) =  0$. Consequently, the second term in Equation \ref{eq:pushright2}, $D_2 \rightarrow 0$ as ${n} \rightarrow \infty$ due to Theorem 1 of \cite{Csorgo1983}. Further, $D_3 \leq \mathcal{V}\left(\bar{A}_{n}\right)/(2\pi)^d$, since $\left|\phi(\vect{t}) \right| \leq 1$. Consequently,  $D_3 \rightarrow 0$ as ${n} \rightarrow \infty$. To prove $D_1 \rightarrow 0$ as ${n} \rightarrow \infty$, we first consider the two following disjoint sets, 
\begin{equation*}
    \begin{aligned}
        B^{+}_{n} &= \{\vect{t}: \lvert \mathcal{C}(\vect{t}) \rvert^2 \geq 4({n}-1)/{n}^2\}, \\
        B^{-}_{n} &= \{\vect{t}: \lvert \mathcal{C}(\vect{t}) \rvert^2 < 4({n}-1)/{n}^2\}. 
    \end{aligned}
\end{equation*} Using Equation \ref{eqn:phi_optim_ap_solved}, we rewrite the first integral $D_1$ as follows
\begin{equation*}
    \begin{aligned}
    D_1  &=  (2 \pi)^{-d}  \int_{A_{n} \cap B^{+}_{n}} \lvert \mathcal{C}(\vect{t}) \rvert 
        \left(1-\frac{{n}}{2({n}-1)}\left[1+\sqrt{1-\frac{4({n}-1)}{ \lvert{n} \mathcal{C}(\vect{t})\rvert^{2}}}\right] d \vect{t} \right)  +  (2 \pi)^{-d}  \int_{A_{n} \cap B^{-}_{n}} \lvert \mathcal{C}(\vect{t}) \rvert d \vect{t} \\ 
        &:= D_4 + D_5.
        \end{aligned}
\end{equation*} The first term $D_4$ may be simplified by noting that for $\vect{t} \in B^{+}_{n}$, we have $\lvert \mathcal{C}(\vect{t}) \rvert^2 \geq 4({n}-1)/{n}^2$. This ensures a non-negative argument under the square root operation. Using the inequality $\sqrt{1-x} + \sqrt{x} \geq 1$ for $0 \leq x \leq 1$ for $D_4$, and using the inequality $\lvert \mathcal{C}(\vect{t}) \rvert \leq \sqrt{4({n}-1)/{n}^2} \text{ for } \vect{t} \in B_{n}^{-},$ we establish that $D_1$ is bounded as follows:
\begin{align}
\label{eq:pushright3}
       D_1 &= D_4 + D_5 \notag\\
       & \leq  (2 \pi)^{-d}  \int_{A_{n} \cap B^{+}_{n}}  \left\{ \frac{1}{\sqrt{{n}-1}} - \frac{\lvert \mathcal{C}(\vect{t}) \rvert}{{n}-1} \right\}   d\vect{t}   
        +  (2 \pi)^{-d} 
        \frac{\sqrt{4({n}-1)}}{{n}} \int_{A_{n} \cap B^{-}_{n}} d \vect{t} \notag\\
        &\leq   (2 \pi)^{-d}  \left\{ \frac{1}{\sqrt{{n}-1}} + \frac{1}{{n}-1} \right\} \int_{A_{n} \cap B^{+}_{n}} d\vect{t}   + 
         (2 \pi)^{-d}  \frac{\sqrt{4({n}-1)}}{{n} } \int_{A_{n}} d \vect{t} \notag\\
        &=  (2 \pi)^{-d}   \left\{ \frac{1}{\sqrt{{n}-1}} + \frac{1}{{n}-1} \right\} \mathcal{V}(A_{n} \cap B_{n}^+) +   (2 \pi)^{-d}  \frac{\sqrt{4({n}-1)}}{{n}} \mathcal{V}(A_{n} \cap B_{n}^-) \notag\\
        & \leq  (2 \pi)^{-d}  \left\{ \frac{1}{\sqrt{{n}-1}} + \frac{1}{{n}-1} + \frac{\sqrt{4({n}-1)}}{{n}} \right\} \mathcal{V}(A_{n}) \notag\\
        & \leq  (2 \pi)^{-d}  \left\{ \frac{1}{\sqrt{{n}-1}} + \frac{1}{{n}-1} + \frac{2}{\sqrt{{n}-1}} \right\}  \mathcal{V}(A_{n}).
        \end{align} 
The assumptions in Theorem \ref{thm:strong_conv1} include $\mathcal{V}(A_{n})/\sqrt{{n}} \rightarrow 0$ as $n \rightarrow \infty$, which ensures that the upper bound in Equation \ref{eq:pushright3} tends to zero for large ${n}$. In summary, assuming  $\mathcal{V}(A_{n}) \rightarrow \infty$, $\mathcal{V}(\bar{A}_{n}) \rightarrow 0$, and $\mathcal{V}(A_{n})/\sqrt{{n}} \rightarrow 0$ as $n \rightarrow \infty$, we have $\left| \hat{f}_{SC}(\vect{z}) -  {f}_{\vect{z}}(\vect{z})\right| \rightarrow 0$ for every $\vect{z} \in \mathcal{Z}$. In fact, we can claim $\sup_{\vect{z} \in \mathcal{Z}}\left| \hat{f}_{SC}(\vect{z}) -  {f}_{\vect{z}}(\vect{z})\right| \rightarrow 0$ as $n \rightarrow \infty$. Further, assuming the true density ${f}_{\vect{z}}$ is continuous on $\mathcal{Z}$ we have uniform strong convergence of $\hat{f}_{SC}$ to $f_{\vect{Z}}$ as $n \rightarrow \infty$.

\subsection{Proof of Theorem \ref{thm:mi-consistency2}: Consistency of $\widehat{MI}_{fast}$.}
\label{subsubsec:proof_thm2}
In this section, we prove Theorem \ref{thm:mi-consistency2} which establishes that the estimator $\widehat{MI}_{fast}$ given by Equation \ref{eq:fastMI} converges in probability to the true $MI$ in Equation \ref{eq:copula-entropy}. For this proof we define the oracle estimator
\begin{equation*}
    MI_{oracle} = n^{-1} \sum_{i = 1}^n \ln \left\{ \frac{ {c}_{\vect{U_X}\vect{U_Y}}(\vect{U}_{\vect{X}i}, \vect{U}_{\vect{Y}i})}{ {c}_{\vect{U_X}}(\vect{U}_{\vect{X}i}) {c}_{\vect{U_Y}}(\vect{U}_{\vect{Y}i})} \right\}. 
\end{equation*} Note that by the law of large numbers $MI_{oracle}$ is consistent for $MI$. From Theorem \ref{thm:strong_conv1} we know that for any small $\epsilon$, there exists a large enough $N$ such that $\sup_{\vect{u} \in (0,1)^d} \left| \hat{c}(\vect{u}) -  {c}(\vect{u})\right| < \epsilon$. Using Taylor series expansion to $\ln \left\{\hat{c}(\vect{u}_i)/{c}(\vect{u}_i) \right\}$, we get 
$\ln \left\{\hat{c}(\vect{u}_i) \right\} =\ln \left\{{c}(\vect{u}_i) \right\} + \frac{\hat{c}(\vect{u}_i) - {c}(\vect{u}_i)}{{c}(\vect{u}_i)} + o(\epsilon),$
  where we ignore the last term. Summing the expression above over $N$ observations, we get the following chain of inequalities:
\begin{equation*}
    \begin{aligned}
     \left|\frac{1}{N} \sum_{i = 1}^N \ln \left\{\hat{c}(\vect{u}_i) \right\} - \frac{1}{N}\sum_{i = 1}^N \ln \left\{{c}(\vect{u}_i) \right\} \right| \notag & \leq \frac{1}{N} \sum_{i = 1}^n \left| \frac{\hat{c}(\vect{u}_i) - {c}(\vect{u}_i)}{{c}(\vect{u}_i)} \right| \leq  \frac{\sup_i \left| \hat{c}(\vect{u}_i) - {c}(\vect{u}_i)\right|}{c_0} \leq \frac{\epsilon}{c_0}, \notag
    \end{aligned} 
\end{equation*} where the true copula density is bounded below by $c_0$. The argument above can be repeated individually for all three of the underlying copula density functions ${c}_{\vect{U_X}\vect{U_Y}}$, ${c}_{\vect{U_X}}$, and ${c}_{\vect{U_Y}}$ and their respective estimates $\hat{c}_{\vect{U_X}\vect{U_Y}}$, $\hat{c}_{\vect{U_X}}$, and $\hat{c}_{\vect{U_Y}}$. Finally, we can write
\begin{equation*}
    \begin{aligned}
     \left|\widehat{MI}_{fast} - MI_{oracle} \right| \notag \leq & \left|\frac{1}{N} \sum_{i = 1}^N \ln \left\{\hat{c}_{\vect{U_X}\vect{U_Y}}(\vect{U}_{\vect{X}i}, \vect{U}_{\vect{Y}i}) \right\} - \frac{1}{N}\sum_{i = 1}^N \ln \left\{{c}_{\vect{U_X}\vect{U_Y}}(\vect{U}_{\vect{X}i}, \vect{U}_{\vect{Y}i}) \right\} \right| + \\
    &\left|\frac{1}{N} \sum_{i = 1}^N \ln \left\{\hat{c}_{\vect{U_X}}(\vect{U}_{\vect{X}i}) \right\} - \frac{1}{N}\sum_{i = 1}^N \ln \left\{{c}_{\vect{U_X}}(\vect{U}_{\vect{X}i}) \right\} \right| + \\
    &\left|\frac{1}{N} \sum_{i = 1}^N \ln \left\{\hat{c}_{\vect{U_Y}}(\vect{U}_{\vect{Y}i}) \right\} - \frac{1}{N}\sum_{i = 1}^N \ln \left\{{c}_{\vect{U_Y}}(\vect{U}_{\vect{Y}i}) \right\} \right| \\
    \leq & \epsilon^{\ast},
     \end{aligned} 
\end{equation*} where the last step follows from the intermediate inequality obtained above. Hence, as $MI_{oracle}$ is consistent for $MI$, we can claim $\widehat{MI}_{fast}$ is consistent for $MI$ as well. 

\section*{Acknowledgments}
The authors are grateful to the Associate Editor and two anonymous reviewers for their insightful and constructive comments that led to an improved article and the improvement of the \textsf{R} package. Song’s research is supported by an NSF Grant (DMS2113564). 

\bibliographystyle{00_style}
\bibliography{02_references}

\end{document}